\documentclass[a4paper,10pt]{article}
\usepackage[centertags]{amsmath}
\usepackage{amsmath}
\usepackage[dutch,british]{babel}
\usepackage{}
\usepackage{amssymb}
\usepackage{amsthm}
\usepackage{newlfont}

\theoremstyle{plain}
  \newtheorem{theorem}{Theorem}[section]

  \newtheorem{remark}[theorem]{Remark}
\theoremstyle{definition}

\theoremstyle{remark}

\numberwithin{equation}{section}

\DeclareMathOperator{\Tr}{Tr}

\newcommand\otimesal{\mathop{\hbox{\raise 1.6 ex
  \hbox{$\scriptscriptstyle\mathrm{al}$}
\kern -0.92 em \hbox{$\otimes$}}}}
\newcommand\oplusal{\mathop{\hbox{\raise 1.6 ex
  \hbox{$\scriptscriptstyle\mathrm{al}$}
\kern -0.92 em \hbox{$\oplus$}}}}
\newcommand\Gammal{\hbox{\raise 1.7 ex
\hbox{$\scriptscriptstyle\mathrm{al}$}\kern -0.50 em $\Gamma$}}
\renewcommand\i{\mathrm{i}}

 \let\be=\beta

\let\ka=\kappa \let\la=\lambda \let\om=\omega 
\let\si=\sigma

 \let\Ga=\Gamma \let\La=\Lambda \let\Om=\Omega

\newcommand{\caB}{{\mathcal B}}

\newcommand{\caH}{{\mathcal H}}

\newcommand{\caJ}{{\mathcal J}}
\newcommand{\caK}{{\mathcal K}}
\newcommand{\caL}{{\mathcal L}}

\newcommand{\caW}{{\mathcal W}}

\newcommand{\bbN}{{\mathbb N}}

\newcommand{\bbP}{{\mathbb P}}

\newcommand{\bbR}{{\mathbb R}}

\newcommand{\opunit}{\text{1}\kern-0.22em\text{l}}

\newcommand{\frh}{{\mathfrak h}}

\newcommand{\e}{{\mathrm e}}

\renewcommand{\d}{{\mathrm d}}
\newcommand{\sys}{{\mathrm S}}
\newcommand{\res}{{\mathrm R}}

\renewcommand{\sp}{\mathrm{sp}}

\newcommand{\beq}{ \begin{equation} }
\newcommand{\eeq}{ \end{equation} }
\newcommand{\bet}{ \begin{theorem} }
\newcommand{\eet}{ \end{theorem} }

 \newcounter{smallarabics}
\newenvironment{arabicenumerate}
{\begin{list}{{\normalfont\textrm{\arabic{smallarabics})}}}
  {\usecounter{smallarabics}\setlength{\itemindent}{0cm}
  \setlength{\leftmargin}{5ex}\setlength{\labelwidth}{4ex}
  \setlength{\topsep}{0.75\parsep}\setlength{\partopsep}{0ex}
   \setlength{\itemsep}{0ex}}}
{\end{list}}

\newcounter{smallroman}

\newcommand{\ben}{\begin{arabicenumerate}}
\newcommand{\een}{\end{arabicenumerate}}

\newcommand{\sfock}{\Ga_{\mathrm{s}}}

\newcommand{\wdr}{ }

\newcommand{\refertofinite}{A}

\newcommand{\str}{ |}

\newcommand{\wo}{}

\begin{document}
\begin{center}
\large{ \bf{ Fluctuations of Quantum Currents\\ \small {and}\\
\large{Unravelings of Master Equations  }}} \\
\vspace{15pt} \normalsize

{\bf Jan Derezi\'{n}ski}\footnote{email: {\tt
jan.derezinski@fuw.edu.pl}
}\\
 Department of Mathematical Methods in Physics \\
Warsaw University\\
Ho\.{z}a 74, 01-494 Warszawa,
Poland \\
\vspace{20pt}

{\bf Wojciech De Roeck}\footnote{email: {\tt
 wojciech.deroeck@fys.kuleuven.be} } \footnote{Postdoctoral fellow of the FWO-Flanders}\\
Instituut voor Theoretische Fysica,\\ K.U.Leuven\\
Celestijnenlaan 200D, 3001 Leuven, Belgium\\ [1mm]
 Mathematics Department \\ Harvard University\\
 Oxford Street 1, 02-138  Cambridge, USA
\\
\vspace{20pt}
 {\bf Christian Maes}\\
Instituut voor Theoretische Fysica,\\  K.U.Leuven\\
Celestijnenlaan 200D, 3001 Leuven, Belgium\\

\end{center}

\vspace{20pt} \footnotesize \noindent {\bf Abstract: } The very
notion of a current fluctuation is problematic in the quantum
context. We study that problem in the context of nonequilibrium
statistical mechanics, both in a microscopic setup and in a
Markovian model. Our answer is based on a rigorous result that
relates the weak coupling limit of  fluctuations of reservoir
{observables} under a global { unitary} evolution
 with the
statistics of the so-called quantum trajectories. These quantum
trajectories are frequently considered in the context of quantum
optics, but they remain useful for more general nonequilibrium
{ systems}.
  In contrast with
{the approaches found in the literature,}
we do not assume that the system is
continuously monitored.  Instead, our starting point is 
a relatively realistic unitary dynamics
 of  the full system

\vspace{5pt} \footnotesize \noindent {\bf KEY WORDS:} weak
coupling limit, quantum stochastic calculus, quantum fluctuations,
\vspace{20pt} \normalsize

\section{Introduction}
Certain aspects in  the combination of nonequilibrium physics with
quantum theory are often more problematic than their counterparts
in  nonequilibrium {\it classical} statistical mechanics.  An
important reason is that in statistical mechanics one often starts
from fluctuation theory and from estimates of statistical
deviations. But in the quantum case, even for {\it equilibrium}
statistical mechanics, there is no standard large deviation theory
(although recently progress was made in
\cite{netocnyredig,reybelletld, hiaimosonytomohiro} and in the nonequilibrium case in
\cite{aboesalem}). For nonequilibrium purposes one wants to go
beyond first order perturbation theory around equilibrium, i.e.
beyond covariance estimates. The question then emerges what
{one   accepts as the definition of }
 (also large) fluctuations of heat, work and
currents.

As it is often the case,
 such questions are more prone to
confusion when working in the quantum formalism. Since recent
developments in nonequilibrium physics have focused on
fluctuations (of entropy production), quantum analogues have been
attempted by many different groups, etc.\
\cite{monnaitasaki,matsuitasaki,deroeckmaes,deroeckmaesfluct,kurchanquantum,mukamel,espositoharbolamukamel,espositomukamel,talknerlutzhanggi}.

Throughout this article, we choose the setup of a system connected
to heat
reservoirs and the fluctuations we study, are fluctuations of the heat currents.\\

In Section \ref{sec: basic question}, we present two possible
approaches to fluctuations in a Hamiltonian setup. These
approaches have appeared repeatedly in the above-quoted articles.
We remark that they are equivalent as far as the mean and the
variance are concerned.

In Section \ref{wc}, we put the Hamiltonian description aside in
favor of an effective model -- {a } quantum master equation derived
in the so-called weak coupling limit.  In the framework of this
master equation, we can again distinguish different approaches.
One of these is the formalism of quantum trajectories, which is
discussed in  Section  \ref{qt}. In Section \ref{sec: connecting},
we combine the Hamiltonian description with the effective model.
 Our main result is contained in
formula \eqref{thm: general}. It states that the fluctuations
calculated by quantum trajectories are limits of the fluctuations
in the Hamiltonian description. This result supplements the
well-known derivation of the master equation, be it its rigorous
form, as in \cite{davies1}, or a more 
{ heuristic}
 derivation, as in
\cite{breuerpetruccione}.

One can further remark  that both the fluctuations in the quantum
trajectory picture \cite{deroeckmaesfluct} and the fluctuations in
our Hamiltonian description satisfy the celebrated
Gallavotti-Cohen fluctuation theorem, see \cite{deroeckbrussel}
for a discussion. However, stressing this point would be
misleading because one does not need the weak-coupling limit to
state the Gallavotti-Cohen fluctuation theorem.

The work is mostly inspired by and based on
\cite{deroeckmaesfluct,derezinskideroeck2}. Section \ref{sec:
math} presents a summary of mathematical
 arguments and states the
main message of this paper as a theorem.

\section{What is a current fluctuation?} \label{sec: basic
question}

\subsection{Question}
Imagine several heat reservoirs $\res_k$, indexed by $k \in K$.
Each reservoir is in thermal equilibrium at inverse temperatures
$\beta_k$ and well separated from the others. All heat reservoirs
are connected with a small system $\sys$ through a coupling term
proportional to a small coupling constant $\la$. Formally, the composite
system is described by a quantum
Hamiltonian \beq
 H_\la=H_\sys + \sum_{k \in K }H_{\res_k} + \la \sum_{k \in K}
 \textrm{H}_{\sys-\res_k},
 \eeq
in which one assumes a clear separation between the reservoir
Hamiltonians $H_{\res_k}$ and the interaction $H_{\sys-\res_k}$.
The dynamics of the full system $\sys + \sum_{k} \res_k$ is given
by the unitary group $\e^{-\i t H_\la}$ on a Hilbert
space of the form ``system  $\otimes$ reservoirs.''  Imagine 
that the
coupling between the system and the reservoirs { starts} at a certain
initial moment.  We take the initial state  represented by a
density matrix $\rho_0$ of the form
 \beq \rho_0 =
{\rho_\sys} \otimes \left[\mathop{\otimes}\limits_{k \in K}
\rho_{k,\beta_k} \right],\eeq where the states $\rho_{k,\beta_k}$
are equilibrium states at $\be_k$ on the $k$'th reservoir, and
${\rho_\sys}$ is an arbitrary density matrix on $\sys$.\\

 We want to ask
how much energy has flown out of/into the different reservoirs
after some time $t$, and how this quantity fluctuates. In the
classical setup, there is no ambiguity as to what that means:
There one has a phase space $X$ for the total system with a
Hamiltonian flow $x \mapsto x_t$ and the object of interest is the
variable (function on $X$) \beq \label{def: classical flow} x \mapsto H_{\res_k}(x_t) -
H_{\res_k}(x), \eeq where $H_{\res_k}$ now also represents the
energy of the reservoir. Usually (but not necessarily), one starts
from an initial distribution $\rho_0$, rather than from a fixed
phase space point $x \in X$, such that the above variables are
actually random variables subject to some overall constraints like
total energy conservation. A natural method  to study the
fluctuations of \eqref{def: classical flow} is to proceed via its characteristic function, for
$\ka \in \bbR^{|K|}$, with coefficients $\ka_k$: \beq
\label{classical fluctuations}
 \int \rho_0(\d x) \; \e^{ -  \i  \sum_{ k \in
K}\ka_k \left(H_{\res_k}(x_t) - H_{\res_k}(x) \right) }. \eeq
Often, this formula is expressed in terms of a time-integrated
current.

 We now switch back to the quantum case. It seems that there is
more than one way to generalize \eqref{classical fluctuations}, as
has been remarked by several authors.\\

\subsection{Answer 1}
 The question amounts to choosing a
quantization { 
of the observable contained in \eqref{classical fluctuations}.}
 Following the { usual}  quantum dictionary one is
tempted to introduce a ``current operator'' \beq \label{cop} I_k
(t):=- \i U_{-t}^\la[H_\la,H_{\res_k}] U_{t}^\la, \eeq where
$U_t^\la:=\e^{-\i t H_\la}$ is the dynamics generated by the total
Hamiltonian  $H_\la$ and $H_{\res_k}$ is the free Hamiltonian of
the $k$'th reservoir only. Obviously, \beq\label{nr} U^\la_{-t}
H_{\res_k}U^\la_{t}-H_{\res_k}= \int_0^t \d u \;I_k(u). \eeq and
one might  set out to study fluctuations of the heat by
considering fluctuations of the operator \eqref{nr}. This amounts
to replacing \eqref{classical fluctuations} directly with \beq
\label{elegant analogue} \rho_0 \left[\e^{-\i \sum_{ k \in
\caK}\ka_k \left( U^\la_{-t} H_{\res_k}U^\la_{t}- H_{\res_k}
\right)}\right]. \eeq The expression (\ref{elegant analogue})
 looks rather elegant but we do not know
of any experiment or  theoretical consideration
where this quantity enters
naturally. 
Observe for example that
$U^\la_{-t} H_{\res_k}U^\la_{t}$ does not in general commute with
$H_{\res_k}$,  hence their difference 
{does not seem to be easily measurable.}
We therefore prefer a more operational definition that we
present in the next section. Nevertheless, the current operator as
defined through \eqref{cop} has its place in the literature,
 e.g.\ in the quantum formulation of the
Green-Kubo relations \cite{weissbook,jaksicogata}. As we remark in
Section \ref{sec: comparison}, the definition \eqref{cop}
coincides in the first and second order with the definition which
we present below. Hence, for the Green-Kubo relation, it does not
matter which definition of  current fluctuations one chooses.

\subsection{Answer 2}
 A  quantity
 that seems
 more relevant practically
is the following: Assume for simplicity that $(H_{\res_k})_{k \in
K}$ have discrete spectrum, indicating that we have not taken the
thermodynamic limit and let $x \in X$ label a complete set of
eigenvectors $| x \rangle$ of $(H_{\res_k})_{k \in K}$ with
 eigenvalues $(H_{\res_k})_{k \in K}(x)$. The
corresponding spectral projections are denoted $P_x:= |x\rangle
\langle x|$ and, by a slight abuse of notation, we use the same
symbol $P_x$ to denote $1 \otimes P_x$, where $1$ is the identity
on $\caH_\sys$. Then we define the characteristic function as \beq
\label{alternative analogue} \chi(\ka,t,\la,\rho_0):= \sum_{x,y
\in X} \Tr \left[ P_y U^\la_{t} P_x \rho_0 P_x U^\la_{-t}
P_y\right]  \e^{- \i \sum_{k \in \caK} \ka_k \left( H_{\res_k}(y)
- H_{\res_k}(x) \right) }.
 \eeq
The idea behind this formula is clear: measure the reservoir energies (thereby projecting
the reservoirs on the eigenstates $x$), { 
at time $s=0$
switch on the interacting time
evolution $U^{\la}_s$,
at time $s=t$ switch the interaction off, and
 finally measure again the reservoir energies (projecting on the
eigenstates
$y$).}

 We now use that the initial state $\rho_0$ is diagonal in
the basis $|x \rangle$ to rewrite \eqref{alternative analogue} as
 \beq
\label{alternative analogue2}\chi(\ka,t,\la,\rho_0)= \rho_0 \left[
 \e^{-\i \sum_{k \in \caK} \ka_k H_{\res_k}   } U^\la_{t}
\e^{\i \sum_{k \in \caK} \ka_k H_{\res_k}   }  U^\la_{-t} \right].
 \eeq

Actually,  we take the
expression  \eqref{alternative analogue2} just as our
starting point. Our main result is valid only after taking thermodynamic limit,
in which the operators
$(H_{\res_k})_{k \in K}$ have continuous spectrum and the definition of the
 state
$\rho_0$ has to be reconsidered. Nevertheless
 we will see further on that \eqref{alternative analogue2}
 has a well defined thermodynamic
limit.

{ Usually, when considering  a system interacting with
  reservoirs, it is assumed that only  system observables can be
  measured, since the reservoirs are very large and difficult to control.
  \eqref{alternative analogue2} does not follow this
  rule: it
involves measuring reservoir observables $ H_{\res_k}$. 
Note, however, that  $ H_{\res_k}$ are reservoir observables of a special
  kind: they commute with $H_0$, and hence they are constants of motion for the
  free dynamics, which acts outside
 the time interval $[0,t]$.
 Measuring of  $ H_{\res_k}$ at times
 $0$ and $t$ is
  conceivable even if the reservoirs are large (but finite), since
to do this we have an infinite amount of time: $s\in]-\infty,0]$
 for the initial
  measurement and $s\in[t,\infty[$ for the final one. Therefore,
 in our opinion, 
 \eqref{alternative analogue2} can be viewed as
 measurable in realistic experiments, even though it involves reservoir
  observables.}

This approach to fluctuations was already used in
\cite{kurchanquantum} for fluctuations of heat, in
\cite{deroeckmaes} and very recently in \cite{talknerlutzhanggi}
for fluctuations of work, and, most widespread, starting in
\cite{lesovikexcess,lesoviklevitovcharge}, for fluctuations of charge
transport. Note also the elegant approach to statistics of charge
transport in \cite{klich,avronbachmann}.

\subsection{Comparison} \label{sec: comparison}

{ As we have seen,}
 the question ``what is a current fluctuation?'' does not
have a unique answer.
In any case, both definitions of current
fluctuations {we have discussed above}
coincide in the first and second moment. In other
words, the first and second derivatives with respect to $\ka$ of
expressions \eqref{elegant analogue} and \eqref{alternative
analogue2} coincide {and are equal to
\beq\rho_0\left[
\int_0^tI_k(u)\d u\right],\ \ \ 
\rho_0\left[
\int_0^t I_k(u)\d u\int_0^t I_{k'}(u')\d u'\right]\label{var}\eeq
or, alternatively, to 
\beq\rho_0\left[
U_{-t}H_{\res_{k}}U_{t}- H_{\res_{k}}\right],\ \ \ 
\rho_0\left[
    (U_{-t}H_{\res_{k}}U_{t}- H_{\res_{k}}) (U_{-t}H_{\res_{k'}}U_{t}- H_{\res_{k'}}) \right].\eeq
This can be easily checked (disregarding possible subtleties due
to  unboundedness of  operators) by using  that $[H_{\res_{k}}, H_{\res_{k'}}]=0$ and that $\rho_0$ is diagonal in $ H_{\res_{k}}$, i.e.\ for all operators $A$:
 \beq
 \rho_0 \left[ H_{\res_k}  A \right] =  \rho_0 \left[  A  H_{\res_{k}}\right] .
  \eeq
  
Conclusion:  If one is interested in second order fluctuations, for example Green-Kubo and Onsager relations, both generating functions  \eqref{elegant analogue} and \eqref{alternative
analogue2} are equivalent.   

Note that the equality between  \eqref{elegant analogue} and \eqref{alternative
analogue2} up to second order depends crucially on the choice of the initial state. 
However, often in physics, one is  interested in expressions which are independent of the initial state. 

Assume that the dynamics $U_{-t}\cdot U_{t}$ admits a unique nonequilibrium steady state $\rho_{\infty}$\footnote{Obviously, one needs the thermodynamical limit for this assumption to be realistic}.  One expects that the covariance
 \beq \label{covariance gk1} \lim_{t \uparrow +\infty} \frac{1}{t}\rho_0 \left[ \int_0^t
 \d u (I_k(u) - \rho_0 \left[  I_k(u)\right]) \int_0^t  \d u'
(I_{k'}(u') - \rho_0 \left[ I_{k'}(u')\right])   \right]
 \eeq
which is a combination (see also \eqref{ldp covariance}) of the two expressions in \eqref{var} for $t \uparrow \infty$,  is independent of the initial state and in particular equal to the correlation function
  \begin{eqnarray}
&&\int_{\bbR} \d u  \, \rho_{\infty} \bigg[
\big(I_k(u)-\rho_{\infty} \left[I_k(0) \right]\big)
\big(I_{k'}(0)-\rho_{\infty} \left[I_{k'}(0) \right]\big) \bigg],
 \label{current current correlation}
\end{eqnarray}
Using the above reasoning and standard manipulations of the cumulant generating function, one easily checks that \eqref{current current correlation} equals
\beq \label{ldp covariance}
 -\lim_{t \uparrow +\infty} \frac{\partial^2}{\partial \ka_{k}  \partial \ka_{k'} }  \frac{1}{t} \log{ \Big[ \textrm{Expression \eqref{elegant analogue} or \eqref{alternative
analogue2} }\Big] } \Big\str_{\ka=0}
\eeq
When $\rho_{\infty}$ is an equilibrium state, hence all temperatures equal, then expression \eqref{current current correlation} with  $\rho_{\infty} \left[I_k(0) \right]=0$ features in the  Green-Kubo relation, as stated  rigorously in \cite{jaksicogata}
and specifically  for quasi-free systems in
 \cite{aschbachertransport}.

 Having chosen a definition of  a current fluctuation
in a microscopic (Hamiltonian) description, we set out to consider
an effective model, arising by a certain consistent approximation.
The model we will be dealing with in the present paper is that of
the quantum master equation which often arises in the so-called
weak coupling limit. We give some reminders in the next section,
and we continue the answer to our question in Section \ref{qt}. In
Appendix B, we outline the weak coupling limit of both generating
functions \eqref{elegant analogue} and \eqref{alternative
analogue2}.

\section{Weak coupling limit}\label{wc}

One of the aims of the present paper is to supplement the user's
manual for one of the best known effective equations in the
physics of open quantum systems: the master equation for the
\wo{long-time} evolution of the density matrix of a small system
\wo{with discrete spectrum} in contact with reservoirs. It is
widely accepted that master equations gives a good description of
the degrees of freedom of a small system in certain limiting
regimes. In the physics literature, this limiting regime is
characterized by the \emph{Born-Markov} approximation and the
\emph{rotating wave} approximation, see
\cite{alickiinvitation,breuerpetruccione} for a review. Another
name, common especially in the mathematical physics literature,
 is  the
\emph{weak coupling limit}, which makes these two approximations
exact.  It goes back to \cite{vanhove} and was made precise in
\cite{davies1}. An interesting review is contained in
\cite{lebowitzspohn1}.
We start with a
formal sketch of the usual set-up.\\

The small system is modeled by a finite-dimensional Hilbert
space $\sys$ and a Hamiltonian $H_\sys$ -- a certain Hermitian matrix.
It interacts with an environment, possibly containing
several reservoirs indexed by $k \in K$. Let us think
of each reservoir as an assembly of free oscillators. We use
the well-known notation
 \beq \label{familiar not1}
 \d \Ga (h_k) = \int_{\bbR^d} \d q  \, h_k(q) a_k^*(q)a_k(q),
 \eeq
where $h_k$ is a \wdr{dispersion function} on $\bbR^d$, the so
called one-particle Hamiltonian on the one-particle Hilbert space
$\frh_k:=L^2(\bbR^d)$ \wo{and $\d \Ga (h_k)$ acts on $\sfock
(\frh_k)$, the bosonic Fock space corresponding to $\frh_k$.}

 The coupling between the system and the environment is
linear in the sense that \beq \label{familiar not2}
H_{\sys-\res_k} = \int_{\bbR^d} \d q  \, V_k \otimes \big[f_k(q)
 a_k^*(q) + \overline{f_k(q)}   a_k(q)\big],
\eeq where $f_k \in \frh_k$  and $V_k$ are self-adjoint operators
on $\caH_\sys$. The total Hamiltonian is formally given by \beq
\label{eq: total hamiltonian} H_\la = H_\sys \otimes 1+ \sum_{k
\in K} 1 \otimes \d \Ga (h_k)+ \la \sum_{k \in K} H_{\sys-\res_k}.
\eeq on the Hilbert space $\caH_\sys \otimes \left[\otimes_{k \in
K} \sfock (\frh_k) \right]$.  Observe that the prefactor $\la$
measures the interaction strength.

The Hamiltonian $H_\la$ generates a quantum evolution
$U^\lambda_\tau = \e^{-\i \tau H_\la}$.  The weak coupling limit
concerns the convergence of the reduced dynamics on the small
system. The coupling $\la\downarrow 0$ gets very weak as the time
$\tau = t/\lambda^2$ goes to infinity. A well-known
theorem by Davies \cite{davies1}  states the following result (which
is written here in a formal way, one actually needs the framework
of Section \ref{sec: math} or a limiting procedure like in
Appendix \refertofinite {} to give it a precise meaning):
\begin{equation}\label{wcl}  \lim_{\la \searrow 0} \rho_0
 \left[   U^{\la}_{\la^{-2 }t} U^{0}_{-\la^{-2} t}
   (S \otimes 1) U^{0}_{\la^{-2 }t}U^{\la}_{-\la^{-2 }t}
   \right] = {\rho_\sys} \left[ \e^{t \caL} S \right]
\end{equation}
for matrices $S$ on $ \caH_\sys$. The initial state $\rho_0 =
{\rho_\sys} \otimes [\otimes_{k \in K} \rho_{k,\beta_k}]$ is the
product of an arbitrary \wo{state (density matrix)} ${\rho_\sys}$
on $ \caH_\sys$ and  of thermal states $\rho_{k,\beta_k}$ at
inverse temperatures $\be_k$ in the respective reservoirs
$\res_k$. The superoperator (acting on matrices) $\caL$ is the
{ generator of a completely positive dynamics obtained
in the weak coupling limit. It can be written in 
 the so called Lindblad form
\cite{lindblad}, see \eqref{def:
generator} below.} There are of course technical conditions for
\eqref{wcl} to be true, which we skip here. One should realize that
\eqref{wcl} is a non-trivial statement. It specifies conditions
under which the (reduced) evolution on the small system $\sys$
gets autonomous. The result is however somewhat rough, at least
for nonequilibrium practice, as the resulting process is just 
a jump
process between energy levels
of the system. { In particular, it does not allow us
to track the interactions with a
given reservoir.}
  What follows is a way
to remedy that, at least on a { phenomenological} level.\\

A first observation \wo{(which can be checked from the explicit
construction given in Sections \ref{qt} and \ref{sec: math})} is
that the generator splits naturally as \beq
 \label{def: decomp}\caL = \sum_{k \in K} \caL_{k},
 \eeq where each $\caL_k$ can be
defined as the object that would emerge in the weak coupling limit
from the microscopic Hamiltonian by cutting the interaction with
all spaces $\frh_{k}$ except for $k'=k$.

Let us see how the decomposition \eqref{def: decomp} can inspire
us further to answer the question of Section 2 in the case of
master equations.
 One can define the time-evolved current operators \beq \label{def: current op
 wc}
J_k (t) = \e^{t \caL} (\caL_k(H_\sys)),\eeq (much in the spirit of
\eqref{cop}), where we recall that $H_\sys$ is the Hamiltonian of
the small system.

Now, one might conjecture that the characteristic function \beq
\label{def: gen function wc} \rho_\sys \left[ \e^{ -\i
\sum_{k}\ka_k \int_0^t \d u J_k (u)} \right] \eeq is the limit of
the correlation function \eqref{elegant analogue} in the weak
coupling limit. That is actually not correct!  Neither is it
correct that
 \eqref{def: gen function wc} is the limit of \eqref{alternative
 analogue2}.   \eqref{def: gen function wc}, { however, coincides}
 with the limits of both \eqref{elegant analogue} and \eqref{alternative
 analogue2} as far as the mean current (first moment, \wo{ i.e.\ first derivative in $\ka$}) is concerned.

Nevertheless, in \cite{lebowitzspohn1}, one starts from the
current operators \eqref{def: current op  wc} and one obtains the
correct Green-Kubo
 relations. This is due to the fact that one does \textbf{not} calculate
 the variance of the current via the characteristic function \eqref{def: gen function
 wc}, \wo{but instead, one starts from the current-current correlation function which is an analogue of \eqref{current current correlation}}.  (See \cite{deroeckmaesfluct} for a more general treatment of the Green-Kubo relations in the weak coupling limit)

 In the next section, we present another (better) way to identify
 the fluctuations in the weak coupling limit.

\section{Quantum trajectories}\label{qt}


Let us look a bit closer at each of the components $\caL_k$ of the
weak coupling generator. They are given by \beq \label{def:
generator}\caL_k(S)= \i [E_{k}, S]  +\sum_{\om \in \sp
([H_\sys,\cdot])}  c(\om,k) \left( V_{\om,k} S V_{\om,k}^* -
\frac{1}{2}\{V_{\om,k} V_{\om,k}^*,S \} \right), \eeq where
$c(\om,k)$ are positive constants, $E_{k}$ are effective
Hamiltonians, sometimes called \emph{Lamb-shifts}, and
 \beq V_{\om,k}:= \sum_{e,e' \in \sp H_\sys, e-e'=\om}
P_e V_k P_{e'}, \eeq where now $P_e$ are spectral projections  of $H_\sys$
 corresponding to the eigenvalue $e$. The summation in
\eqref{def: generator} is over all differences of eigenvalues of
$H_\sys$ (or equivalently, over all eigenvalues of $[H_\sys,\cdot]$) and in what follows we write simply $\sum_{\om}$ for
$\sum_{\om \in \sp ([H_\sys,\cdot])}$ and $\sum_k$ for $\sum_{k
\in K}$.
 One now decomposes $\caL_k= \caL_{k}^0+\sum_\om \caJ_{\om,k}$ with
\beq \caL_k^0(S)= \i [E_{k}, S] - \frac{1}{2}\sum_{\om} c(\om,k)
\{V_{\om,k} V_{\om,k}^*,S \}, \qquad \caJ_{\om,k} (S)=  c(\om,k)
V_{\om,k} S V_{\om,k}^*,   \eeq where $\caJ_{\om,k}$ is called a
jump operator.
  It is important to keep in mind that such a splitting $\caL_k= \caL_{k}^0+
\caJ_{\om,k}$ is not uniquely given by the generator $\caL_k$,
instead we have used information about the operators $H_\sys$ and
$V_k$ to define this splitting. (See \cite{breuerpetruccione} for
extensive comments on this non-uniqueness).

  The final \emph{
unraveling } is written as
 \beq \label{eq:
splitting} \caL= \caL_0 + \sum_{\om,k} \caJ_{\om,k} ,   \eeq where
$\caL_0=\sum_k \caL_{k}^0$. We now introduce completely positive
operations $\caW_t(\si)$ and $\La^0_t$ given by
\begin{equation} \label{explicitmap}
\mathcal{W}_t(\sigma):= \La^0_{t_1}  \mathcal{J}_{\om_1, k_1}
\La^0_{t_2-t_1}  \ldots \La^0_{t_n-t_{n-1}}
\mathcal{J}_{\om_n,k_n} \La^0_{t-t_n} ,\qquad \La^0_t= \e^{t
\caL_0},
\end{equation}
for a ``trajectory'' $\sigma$ that labels all the jump times and
actions in $\mathcal{W}_t(\si)$: \beq \label{def: si} \sigma =
(t_1,k_1,\om_1 ;t_2,k_2,\om_2;\ldots; t_n,k_n,\om_n),\quad 0\leq
t_1\leq \ldots \leq t_n\leq t. \eeq Via the (norm convergent) Dyson
expansion corresponding to the splitting \eqref{eq: splitting}, we
have
 \beq \label{eq:
dyson expansion} \e^{t \caL } = \int\d \sigma\,\caW_t (\si), \eeq
where the integral over $\sigma$ is the  abbreviation of the following
expression:
\[\sum_{n=0}^{\infty} \sum_{k_1,\ldots,k_n } \sum_{\om_1,\ldots,\om_n }
\int_{0}^{t} \d t_n \int_{0}^{t_n} \d t_{n-1} \ldots
\int_{0}^{t_2} \d t_1 \, =: \, \int\, \d \si.
\]

The  idea is now to interpret each contribution to the sum and
integral as a \emph{ quantum trajectory}. In that philosophy, the
map \beq {S} \mapsto \caW_t(\si) ({S})
 \eeq
 gives the (unnormalized)  evolution of the system, \emph{conditioned} on
 the trajectory $\si$. This conditioning usually means that
 certain measurement outcomes were obtained and that these
 outcomes are represented by $\si$.

Taking this idea just one step further, one can obtain statistics
of measurement outcomes (or, in our case, currents). Define the
following probability distribution on all possible $\si$:
 \beq\label{cd}  \d \bbP^t_{\rho_\sys} (\si) :=  \Tr [{\rho_\sys} \caW_t(\si)(1)] \,\d
\si, \eeq and introduce the energy counting numbers
 $n^t_k$ \beq
n_k^t(\si) := \sum_{i}  \delta_{ k_i,k } (\si) 1_{(t_i <t)}(\si)
\om_i(\si),  \eeq where the index $i$ runs over all jumps present
in $\si$ (i.e.\ from $1$ to $n$ in \eqref{def: si}) and $1_{(t_i
<t)}$ is the indicator function of the event that in $\si$, the
$i$'th jump occurs before time $t$. The distribution of the random
variables $n_k^t$ is inherited from \eqref{cd} and will be used
throughout. The characteristic function of the joint
 distribution on $n^{t}_{k \in K}$ is defined as
 \beq
\label{eq: fluct unravel} \chi_{\mathrm{w.c.}} (\ka,t, {\rho_\sys}
):= \int \d \bbP^t_{\rho_\sys} (\si) \left[ \e^{-\i \sum_{k}
\ka_{k} n^{t}_{k}(\si) } \right]. \eeq Note that  \eqref{eq:
fluct unravel}  characterizes the full distribution of  $n^{t}_{k \in K}$.
  The study of fluctuations indeed amounts to more than characterizing covariances, as useful
  as that may be within linear response theory.

A further point concerns the ``classical'' nature of the variables
$n^t_k(\sigma)$ and the use of standard probability theory. But
that is exactly the point of the present paper: these variables
characterize the ``quantum'' fluctuations, see further in
\eqref{thm: general}.

 In the quantum optics literature, this procedure of ``unraveling
 master equations into trajectories"
is generally accepted, see e.g. \cite{carmichael},
\cite{breuerpetruccione} and the recent review
\cite{belavkinreview}, whereas the first source of unraveling is
probably in \cite{srinivasdavies}.

 From a more fundamental
point of view, one could ask how quantum trajectories and, more
specifically, the variables $n_k^t$ emerge from microscopic
dynamics.
{ The usual justification of quantum unravelings
found in the literature supposes that
  the system is described by a so-called quantum Langevin dynamics (the
  solution of a  quantum stochastic differential equation (QSDE)). 
Even though it is a unitary
  dynamics, it can be viewed only as a very approximate description of
  realistic quantum systems quite far from ``first
  principles''. 
To justify the use of QSDE it is usually assumed that the
unitary evolution is interrupted
 by measurements, which, in the limit of very short times between measurements,
 yields a quantum Langevin
 equation
 (originally introduced by \cite{hudsonparathasaraty} in a mathematical framework, see \cite{attalreview} for recent developments and \cite{gardinerzoller} for a physical point of view.)
  In many cases it amounts to supplementing the quantum
 evolution of the small system  with a
stochastic evolution of classical variables in the environment.
 
The  quantum stochastic differential 
equation (QSDE) or stochastic Schr{\"o}dinger evolutions, and their
 solutions,
  have been
 studied by many authors.  The random variables $n_k^t$ correspond there to
 the fluctuations of (linear combinations of) number operators in the
 reservoir spaces of the QSDE. (see e.g.\ \cite{deroeckmaesfluct}
 or \cite{barchielli})
 We will not elaborate on this point, since it is not our central
 subject (See however Appendix B).

 Lately, a new class of models was considered, which give
 also rise to QSDE's. These are the so-called ``repeated
 interaction models", where, instead of measuring the reservoir
 continuously, one refreshes it, see
 \cite{attalpautrat,attaljoye}.
}

\section{Connecting unravelings with the Hamiltonian
dynamics}\label{sec: connecting}

{ Our paper describes an alternative justification of quantum unravellings that
we believe is in many situations more satisfactory
 and does not explicitly  introduce continual
 measurements, stochasticity or ``refreshing of reservoirs''. We start from a
class of dynamics described in Section  \ref{wc}, which  are often viewed
as a relatively adequate description of realistic quantum
systems. We prove  that, after first 
applying the thermodynamic limit and then taking the
weak coupling limit, the quantities \eqref{alternative analogue2} converge 
to quantum unravellings --expressions of the form
\eqref{eq: fluct unravel}.
 This idea, to our knowledge, is present in the
literature only in a heuristic form. What we describe is a rigorous result.

Our proof of this result involves two steps. The first step is the theorem
about the extended weak coupling limit obtained by two of us
\cite{derezinskideroeck2}, which says that the microscopic 
dynamics converges in a certain
sense to an appropriate
 quantum Langevin dynamics. The second step (which is well
known in the literature) goes from the quantum Langevin dynamics to quantum
unravellings. 

There exists actually an alternative proof of our result that goes 
directly from the
microscopic dynamics to quantum unravellings, without passing through a
quantum Langevin dynamics. We will indicate it briefly in Remark
\ref{remarko}. }

Recall that \[  H_{\res_k}=\d \Ga
 (h_{k}) = \int_{\bbR^d} \d q  \, h_k(q) a^*_k(q)a_k(q)\]
 is the (second quantized) $k-$th
reservoir Hamiltonian, cfr. \eqref{familiar not1}.  Remember also
the characteristic function $ \chi_{\mathrm{w.c.}} (\ka,t,
{\rho_\sys} )$ from \eqref{eq: fluct unravel}. Here comes the main
result of the paper.

 Under standard
conditions of the weak coupling limit as in \eqref{wcl}, and with
the same remark about precision as in \eqref{wcl}, \beq\label{thm:
general} \chi_{\mathrm{w.c.}} (\ka,t, {\rho_\sys} )= \lim_{\la
\searrow 0} \rho_{0}
 \left[ \e^{-\i \sum_k \ka_k \d \Ga
 (h_{k})} U^{\la}_{\la^{-2} t}  \e^{\i \sum_k \ka_k \d \Ga
 (h_{k})} U^{\la}_{-\la^{-2} t}
 \right].
\eeq The right-hand side is of course a general instance of the
weak coupling limit of \eqref{alternative analogue2}.  This proves
that quantum trajectories provide nonequilibrium fluctuations of
the time-integrated heat dissipated in a given reservoir. In
particular, the properties of the distribution of $n^t_k$, such as
these related to fluctuation theories or to Green-Kubo relations,
are related to the quantum fluctuations of  energy currents \wo{in
the sense of \eqref{thm: general} and the arguments of Section
\ref{sec: basic question}.} An extensive study of the distribution
associated to \eqref{def: gen function wc} was performed in
\cite{deroeckmaesfluct}.

\section{Mathematical statement of the results}\label{sec:
math}

The main result of the paper, formula  \eqref{thm: general},  is a
consequence of the more general and abstract results proven in
\cite{derezinskideroeck2}. For the convenience of the reader, we
list some simple assumptions which allow to establish \eqref{thm:
general} and we specify what is the exact definition of the
quantities appearing on the RHS \eqref{thm: general}. For simplicity of
presentation,
we choose a
special form for the one-particle Hamiltonians $h_k$ and we
restrict the physical dimension of the reservoir space $d$ to
$d=1$.

\subsection{Formal Hamiltonian}
\label{Formal Hamiltonian}

Recall the formal Hamiltonian from \eqref{eq: total hamiltonian}:
\beq H_\la = H_\sys + \sum_{k \in K}  H_{\res_k} + \la  V_k
\otimes(  a_k^*(f_k)+   a_k(f_k) ),\label{zero} \eeq where now
\begin{itemize} \item $H_\sys=H^*_\sys, V_k=V_k^* \in \caB(\caH_\sys)$
with $\dim \caH_\sys < \infty$;
\item $H_{\res_k}=\d \Ga (h_{{k}})$, where
 $h_{k}$ are the one-particle Hamiltonians on the Hilbert spaces
$\frh_k = L^2(\bbR^+)$  acting as \beq (h_{k}g)(x)= x g(x); \eeq
\item $f_k \in \frh_k$  are coupling functions;
\item $a^*_k(f_k) / a_k(f_k)$ are
 creation/annihilation operators on the bosonic Fock space
 $\Gamma_{\mathrm{s}} (\frh_k)$;
\end{itemize}

\subsection{Effective master equation}

{ Given the information of Section \ref{Formal Hamiltonian}
and the inverse temperatures $\beta_k$, $k\in K$,} one can  construct the
weak-coupling generator $\caL$, which was introduced in Section
\ref{wc} with unspecified
 parameters  $c(\om,k)$ and $E_{k}$.
Define the functions \beq f^{\be_k}_k (x):=\left\{
\begin{array}{ll}
\frac{f_k(x)}{\sqrt{\e^{\be_k x}-1}},&x>0,\\[2ex]
\frac{\overline{f_k(-x)}}{\sqrt{1-\e^{\be_k
x}}},&x<0.\end{array}\right.    \eeq
 The exact expressions for the parameters  $c(\om,k)$ and $E_{k}$
 are (see e.g. \cite{derezinskideroeck2})
\begin{eqnarray}
 c(\om,k)&=&  \frac{1}{2\pi}|f^{\be_k}_k(\om)|^2,  \label{def: rates} \\
E_{k}  &=& \sum_{\om}  V^*_{\om,k}  V_{\om,k} \, \Im \int_{\bbR^+}
\d t \,  \e^{-\i \om t} \int_{\bbR} \d x \,  \e^{-\i
 t x }|f^{\be_k}_k(x)|^2   .  \label{def: effective}
\end{eqnarray}

\subsection{Coupling to thermal reservoirs}

One of the subtle points of quantum statistical physics is how to
describe infinitely extended bosonic reservoirs at positive
temperatures. Strictly speaking, the Hamiltonian $H_\la$ defined
in \eqref{zero} describes the reservoirs only at zero temperature
-- but we are interested in the case of an arbitrary temperature.
 (This is why $H_\la$ was called the ``formal
Hamiltonian''). \wo{We need the state $\rho_0$ in expressions
\eqref{wcl} and \eqref{thm: general} to be a thermal state (which
cannot be represented by a density matrix in infinite volume). In
expression (\ref{wcl}, \ref{thm: general}), we pretended that this
state can be defined on
$\caB(\caH_\sys \otimes [\otimes_{k \in K} \sfock (\frh_k) ])$,
 or at least on its sufficiently large subalgebra preserved by the
  dynamics,
but this could be problematic. (This would however be a good
approach for fermions!) }

Fortunately, there exists a formalism that allows us to describe a
system interacting with reservoirs at positive temperatures in the
thermodynamic limit rigorously. This formalism  was used in
standard works like
\cite{jaksicpillet3,jaksicpillet4,bachfrohlichreturn}.
 The relevance of
this construction has been argued e.g. in
 \cite{derezinski1,jaksicpilletderezinski}. Here we
just present how one should modify the dynamics of the coupled
system \wo{and we point to Appendix \refertofinite{} for a
justification.}

One of the ingredients of this formalism are the so-called Araki-Woods
representations of the CCR. To introduce them one needs to enlarge
the Hilbert space. The enlarged Hilbert space is \beq \caH :=
\caH_\sys \otimes \sfock(\oplus_{k \in K} (\frh_k \oplus \frh_k)
). \eeq
%

We define the  \emph{free Liouvillian} of the $k$'th reservoir on
$\sfock (\frh_k\oplus \frh_k)  $ as  \beq L_{\res_k} = \d \Ga (h_k
\oplus (-h_k)).
 \eeq

From now on, it will be convenient to identify $\frh_k\oplus
\frh_k$ with $L^2(\bbR)$ such that the one-particle operator $h_k
\oplus (-h_k)$ acts by multiplication with $x \in \bbR$.
 The generator of
the dynamics is chosen as the so-called \emph{semi-Liouvillian}
 (see e.g. \cite{derezinskijaksicjsp} for explanations
about the terminology) and it equals \beq L_\la = H_\sys + \la
\sum_{k\in K} V_k \otimes \left( a_k^*(f^{\be_k}_k )
+a_k(f^{\be_k}_k ) \right) +\sum_{k \in K} L_{\res_k}. \eeq
 This is a formal expression, but
one can easily construct the operator $L_\la$ rigorously (see
\cite{derezinskideroeck2}). Finally, let $\Om$ stand for the
vacuum vector in $\sfock(\oplus_{k \in K}( \frh_k \oplus \frh_k)
)$ and define the vacuum state \beq \mathrm{Vac} [\cdot] = \langle
\Om, \cdot \, \Om \rangle. \eeq

We can now define our object of interest: \beq \label{object of
interest gns}\chi (\ka,t,\la,\rho_\sys) = (\rho_\sys \otimes
\mathrm{Vac} ) \left[ \e^{-\i\sum_{k \in K} \ka_k L_{\res_k}}
\e^{-\i t L_\la}
   \e^{\i \sum_{k \in K} \ka_k L_{\res_k}}  \e^{\i
t L_\la} \right].
 \eeq
Remark that \eqref{object of interest gns} arises  from
\eqref{alternative analogue2}, that is
 \beq
\label{operator not in algebra}\chi(\ka,t,\la,\rho_0)= \rho_0
\left[ \e^{-\i \sum_{k \in \caK} \ka_k H_{\res_k}   } U^\la_{t}
\e^{\i \sum_{k \in \caK} \ka_k H_{\res_k}   } U^\la_{-t}
 \right],
 \eeq
{  by replacing $ \left[\mathop{\otimes}\limits_{k \in K}
\rho_{k,\beta_k} \right]$ by $ \mathrm{Vac} $,  $H_{\res_k}$ by
$L_{\res_k}$ and $H_\la$ by $L_\la$. One can check that in finite volume both
 expressions coincide -- in particular, 
the positive temperatures of the reservoirs
 have been incorporated directly into the
functions $f^{\be_k}_k$.} In fact, the vacuum  state $\mathrm{Vac}$
represents  the product of thermal states on the appropriate
algebra of observables. More details and explanations can be found
e.g.\ in \cite{derezinski1}.

In Appendix \refertofinite,  a limiting procedure which
constructs expression \eqref{object of interest gns} via
finite-volume approximations is explained.

\subsection{Main result}

We are ready to state rigorously our main result
\begin{theorem} \label{thm: main}
Let  $\chi_{\mathrm{w.c.}} (\ka,t,  {\rho_\sys} )$ be defined via
 \eqref{eq: fluct unravel} as in Section \ref{qt} and with parameters (\ref{def: rates},\ref{def: effective}). Assume that \ben
\item{The reservoir correlation functions are
integrable \beq  \int_{\bbR^+} \d t \Big| \int_{\bbR} \d x
|f^{\be_k}_k(x)|^2 \e^{-\i t x } \Big|  < \infty. \eeq }

\item{ The functions
$f^{\be_k}_k $ are continuous
 in a neighbourhood of $(\sp H_\sys-\sp H_\sys) $.  } \een

Then for all $\ka \in \bbR^{|K|}$, \beq \lim_{\la \downarrow 0}
\chi(\ka,\la^{-2}t,\la,\rho_\sys) =
\chi_{\mathrm{w.c.}}(\ka,t,\rho_\sys). \eeq
\end{theorem}

This theorem is a simple consequence of Theorem 5.7 in
\cite{derezinskideroeck2}. Essentially, one proves that $\e^{-\i t
\la^{-2}L_\la }$ converges to a quantum Langevin dynamics
 and operators like $\e^{-\i \ka_k
\la^{-2} L_{\res_k}}$ converge to $\e^{-\i \ka_k N_k}$, where
$N_k$ is an appropriate linear combination of number operators
(see also Appendix B). The link with unravelings of master
equations belongs then to standard knowledge in quantum stochastic
calculus (e.g.\ to be found in various forms in
\cite{deroeckmaesfluct,boutenkummerer,barchielli}).

\begin{remark}
The assumptions contain a mild infrared {regularity requirement}, 
since the
assumption about $f^{\be_k}_k$ implies that 
$x \mapsto x^{-1/2}f_k(x) \in L^2(\bbR^+)$.
\end{remark}

\begin{remark}
 Note that one would like to
strengthen the statement of Theorem \ref{thm: main}
For example, one
would like to have convergence of the derivatives in $\ka$. This
is possible under stronger regularity assumptions on $f_k$.
One can also prove a version of   Theorem \ref{thm: main} allowing for complex
 $\kappa_k$, see  \cite{deroeck}.
\end{remark}

\begin{remark}\label{remarko}
Although the extended weak coupling limit in
\cite{derezinskideroeck2} gives valuable insight into the limit of
fluctuations (see Appendix B), one does not really need it to
prove Theorem \ref{thm: main}.  One can also rewrite
$\chi(\ka,t,\la,\rho_0)$ as \beq\label{halfkappageneratingfunction}
 \chi(\ka,t,\la,\rho_0)=
(\rho_\sys \otimes \mathrm{Vac} ) \left[  \e^{-\i t L_{\la,\ka/2}}
    \e^{\i
t L_{\la,-\overline{\ka}/2}} \right]. \eeq where \begin{eqnarray}
 L_{\la,\ka}&=&
H_\sys   +\sum_{k \in K} L_{\res_k}\\
&&+\la \sum_{k\in K} V_k \otimes \left( a_k^*( \e^{-\i \ka_k
(h_k \oplus -h_k)}f^{\be_k}_k ) +a_k( \e^{-\i \overline{\ka}_k (h_k
\oplus -h_k)}f^{\be_k}_k ) \right),\nonumber
\end{eqnarray} which, at least for $\Im \ka =0$, reduces technically to the
usual derivation of the master equation.
\end{remark}

\renewcommand{\theequation}{A-\arabic{equation}}
  \setcounter{equation}{0}  

  \section*{APPENDIX A}  
  \label{app: finite}

We justify \eqref{object of interest gns} from the physical point
of view. We do that by arguing that \eqref{object of interest gns}
is the thermodynamic limit of finite volume versions of
\eqref{operator not in algebra}.

Let for each $n \in \bbN $, $\frh_{k,n}$ be finite-dimensional
Hilbert spaces with one-particle Hamiltonians $h_{k,n}$ and
coupling functions $f_{k,n} \in \frh_{k,n} $. Define the finite
volume evolution reservoir Hamiltonians $H_{\res_k,n}$ and full
Hamiltonian $H_{\la,n}$ by
\begin{eqnarray*}
H_{\res_k,n} &=& \d \Ga (h_{k,n}) \\
H_{\la,n} &=& H_\sys + \sum_{k \in K} H_{\res_k,n}  +\la\sum_{k
\in K} V_k \otimes  \left( a^*(f_{k,n}) +  a(f_{k,n})
\right)\end{eqnarray*} and the finite volume thermal states on
$\caB(\sfock(\frh_{k,n}))$ as \beq \rho_{k,\be_k,n}\left[\cdot
\right] = \frac{1}{Z_{k,n}(\be_k)}\Tr \left[ \e^{-\be_k
H_{\res_k,n} }\, \cdot \right],   \qquad Z_{k,n}(\be_k)= \Tr
\left[ \e^{-\be_k H_{\res_k,n} } \right] \eeq

Assume that \ben
\item{There is a  $C<\infty$ such that for all $t \in \bbR, k  \in K$ and $n \in \bbN$,
\begin{eqnarray*} \label{ass: finite1a}
 &\left| \rho_{k,\be_k,n}\left[ a^*( \e^{-\i t h_{k,n}}f_{k,n}) a( f_{k,n})\right]
\right|&  \leq C
\end{eqnarray*} }
\item{ For each $t \in \bbR$ and $k  \in K$,
\begin{eqnarray*} \label{ass: finite2a} &&\mathop{\lim}\limits_{n\uparrow \infty} \rho_{k,\be_k,n}\left[ a^*( \e^{-\i t h_{k,n}}f_{k,n}) a( f_{k,n})\right]
= \langle f_k, (\e^{\be_k
h_{k}}-1)^{-1}\e^{-\i t h_{k}} f_k \rangle_{\frh_k} \\
\label{ass: finite2b} && \mathop{\lim}\limits_{n\uparrow \infty}
\rho_{k,\be_k,n}\left[ a( f_{k,n}) a^*(\e^{-\i t h_{k,n}}
f_{k,n})\right]=\langle f_k, (1-\e^{-\be_k h_{k}})^{-1}\e^{-\i t
h_{k}} f_k \rangle_{\frh_k}
\end{eqnarray*}
The notation on the RHS was introduced in Section \ref{wc}. }
 \een

 Then the expressions
 \beq \label{def: sequence finite}
\chi_n( \ka,t, \la, \rho_\sys):=  (\rho_\sys
\otimes[\mathop{\otimes}\limits_{k\in K} \rho_{k,\be_k,n}]) \left[
\e^{-\i \sum_{k} \ka_k H_{\res_k,n} } \e^{-\i t H_{\la,n} } \e^{\i
\sum_{k} \ka_k H_{\res_k,n}} \e^{\i t H_{\la,n} }
 \right]
 \eeq
converge  as $n \uparrow \infty$ for all $t$ and $\la$. This is
checked by writing a Dyson expansion for \eqref{def: sequence
finite}, treating the terms in $\la$ as a perturbation.  The
dominated convergence theorem can be applied  since term-by-term
convergence is implied by Assumption (2) above and a dominating
bound follows from Assumption (1) above.

The connection with the setup in Section \ref{sec: math} is that,
under the above assumptions \beq \lim_{n\uparrow \infty}\chi_n(
\ka,t, \la, \rho_\sys) = \chi (\ka,t,\la,\rho_\sys)  \eeq

\renewcommand{\theequation}{B-\arabic{equation}}
  \setcounter{equation}{0}  

 \section*{APPENDIX B}  
  \label{app: qsde}

In this appendix, we look more closely at what the "extended weak
coupling limit" (as presented extensively in
\cite{derezinskideroeck2}) can tell us about the different
fluctuation formulas that were proposed in Section \ref{sec: basic
question}.

Assume the notation introduced in Sections 2 and 3. The result in
\cite{derezinskideroeck2} states that the unitary dynamics
$U_{\la^{-2} t}^\la$ on $\caH_\sys \otimes_{k } \sfock(\frh_k) $
 converges (in an appropriate sense) to a
new unitary dynamics $\tilde{U}_{ t}$ on $\caH_\sys
\otimes_{k,\om} \sfock(\tilde{\frh}_{k,\om}) $, where $\tilde{\frh}_{k,\om}$ are modified one-particle spaces. The dynamics $\tilde{U}_{ t}$, which is the
solution of a Quantum Stochastic Differential Equation, is
extensively discussed in
\cite{deroeckmaesfluct,derezinskideroeck2}.

An important observation is the emergence of  effective reservoir
energy operators; \beq N_k := \sum_{\om} \d \Ga (\om 1_{\om,k})
\eeq where $1_{\om,k}$ is the projector on $\tilde{\frh}_{k,\om}$.
One sees hence that the effective reservoir energy operator is
more like a number operator. Its plays a role in the limits of
respectively \eqref{elegant analogue} and \eqref{alternative
analogue2}. One has, of course again under technical conditions,
\beq  \label{correspondance wc 1}  \rho_0 \left[\e^{-\i \sum_{ k
\in \caK}\ka_k \left( U^\la_{-t} H_{\res_k}U^\la_{t}- H_{\res_k}
\right)}\right] \, \mathop{\longrightarrow}\limits_{\la \searrow 0}  \, \tilde{\rho}_0
\left[ \e^{-\i \sum_{k \in K}\ka_k ( \tilde{U}_{-t} N_k \tilde{U}_{-t}- N_k
) } \right] \eeq
 and
\beq \label{correspondance wc 2} \rho_0 \left[
 \e^{-\i \sum_{k \in \caK} \ka_k H_{\res_k}   } U^\la_{t}
\e^{\i \sum_{k \in \caK} \ka_k H_{\res_k}   }  U^\la_{-t} \right]
\, \mathop{\longrightarrow}\limits_{\la \searrow 0}  \,  \tilde{\rho}_0 \left[  \e^{-\i
\sum_{k \in K}\ka_k  N_k} \tilde{U}_{t} \e^{\i \sum_{k \in K} \ka_k N_k
   } \tilde{U}_{-t}
    \right] \eeq
where $\tilde{\rho}_0$ is a state which coincides with $\rho_0$ on $\caH_\sys$ and which represents the thermal reservoir states on $\otimes_{k } \sfock(\frh_k)$.
Although it is not obvious from these expressions, the expression
\eqref{correspondance wc 2} coincides with \eqref{eq: fluct
unravel}, this is the well-known connection between unravelings
and quantum stochastic differential equations.  To check that
\eqref{correspondance wc 1} and \eqref{correspondance wc 2}
coincide up to second order in $\ka$, it suffices to know that
\beq \tilde{\rho}_0[ A N_k]= \tilde{\rho}_0[ N_k A]= 0 \eeq for
all $A$, operators on $\caH_\sys \otimes_{k,\om}
\sfock(\tilde{\frh}_{k,\om})$.

\section*{Acknowledgments}
W.D.R acknowledges inspiring discussions with K. Neto{\v{c}}n\'y, C-A. Pillet and G.M. Graf.  C.M. benefits from the Belgian Interuniversity
Attraction Poles Program P6/02. W.D.R acknowledges the financial
support of the FWO-Flanders.

\bibliographystyle{plain}
\bibliography{mylibrary07specialforjsp2}

\end{document}